\documentclass[twoside]{ilcws08}
\usepackage[latin1]{inputenc}
\usepackage[dvips]{graphicx,epsfig,color}
\usepackage{wrapfig,rotating}
\usepackage{amssymb,amsmath,array}
\usepackage{subfig}

\begin{document}
\title{
Analysis of the Decay $e^{+} e^{-} \rightarrow \text{ invisible } + H(\rightarrow \mu \mu)$ at a Collision Energy of 500 GeV }  
\author{Jan Strube and Marcel Stanitzki
\vspace{.3cm}\\
Rutherford Appleton Laboratory - PPD \\
Didcot, Oxfordshire OX110QX - UK
\vspace{.1cm}\\
}
\maketitle

\begin{abstract}
The analysis of $e^{+} e^{-} \rightarrow \text{ invisible } + H(\rightarrow \mu \mu)$ at a next generation linear collider presents an opportunity to study the coupling of the Yukawa couplings of the second generation in a clean environment. We give an overview over the experimental challenges of this analysis at a collision energy of 500 GeV and present an outlook to the results of the analysis at a collision energy of 250 GeV.
\end{abstract}

\section{Motivation}
The environment at the ILC is primed for analyses of precision measurements of Higgs couplings. The analysis of the decay $e^{+} e^{-} \rightarrow  \text{ invisible }+H\rightarrow \mu \mu$ is an ideal opportunity to study the Yukawa couplings of the second generation in a clean environment and complement the recoil method measurement of this coupling by taking advantage of the large branching fraction of the decay of the Z boson to neutrinos. In addition to the physics interest, the decay of $H\rightarrow \mu\mu$ in an otherwise (almost) empty detector permits precision studies of the tracking performance and is used for detector benchmarking.

\section{Software}
In absence of fully reconstructed events this analysis was carried out on events generated by Pythia 6.4\cite{pythia} and using the fast detector simulation in the org.lcsim framework. Events were simulated with the sid01 version of the SiD detector concept\cite{sid}. For the classification of events we used the TMVA\cite{tmva} libraries that provide a variety of multi-variate classifiers.

\section{Presentation of Samples}
\begin{figure}
    \subfloat[t channel]{\includegraphics[width=.38\columnwidth]{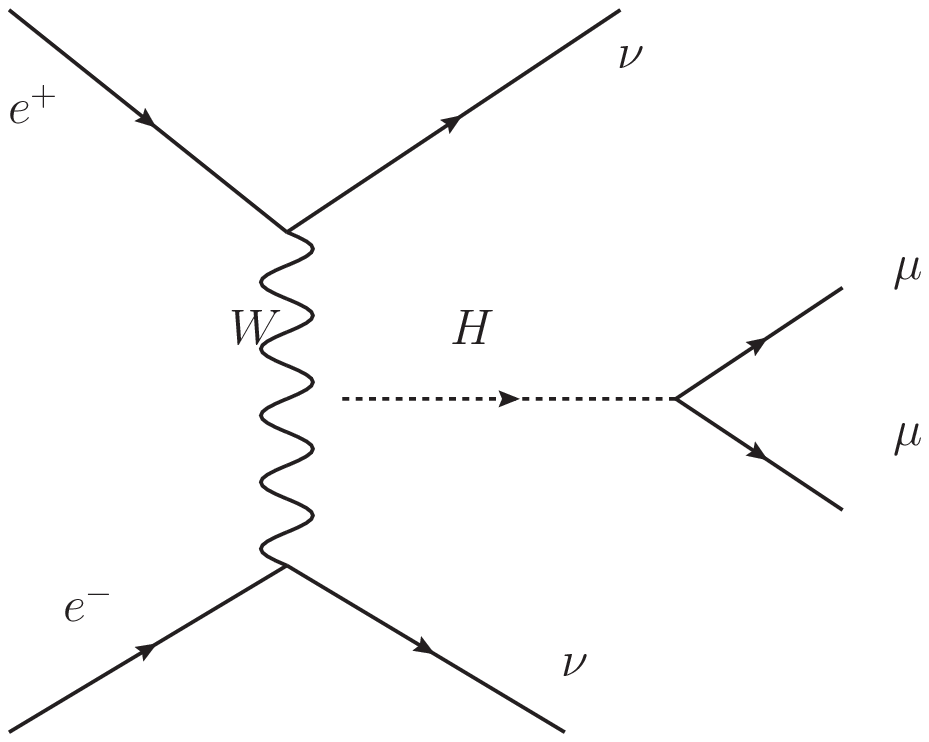}\label{Fig:WH}}
    \hfill
    \subfloat[s channel]{\includegraphics[width=.55\columnwidth]{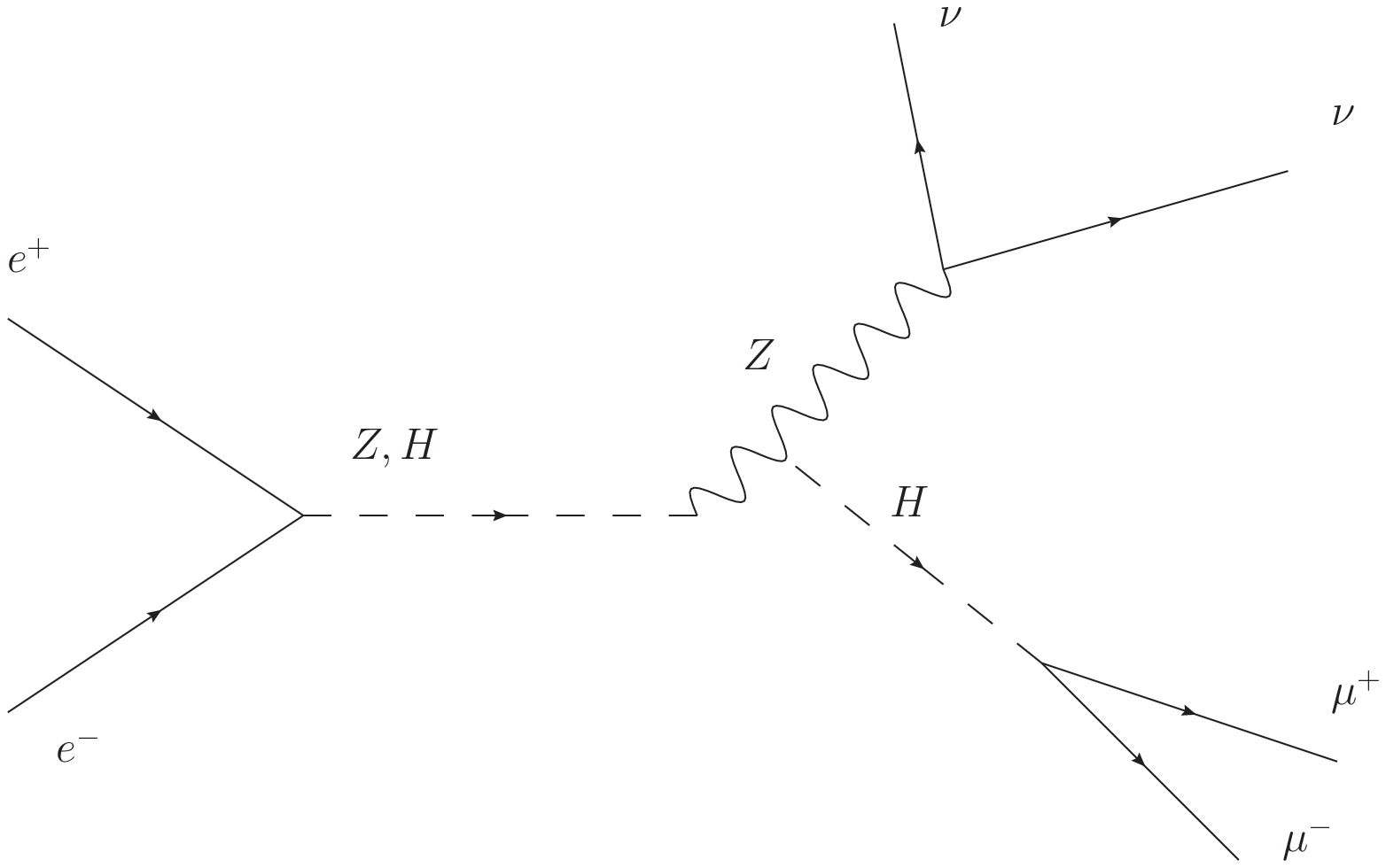}\label{Fig:ZH}}
    \caption{Feynman diagrams of the leading-order signal processes}
    \label{fig:signal}
\end{figure}
Events in the signal sample are computed with either of the two diagrams in Figure~\ref{fig:signal}. The fraction of events in the signal sample containing a Z boson is 16.2\%, the other 83.8\% contain neutrinos from the vector boson fusion diagram\footnote{The relative fraction of the two contributing diagrams to the observed final state are obtained from PYTHIA 6.4}. A mixture of different decays containing a pair of muons in the final state make up the background sample. The different samples and their cross-sections at an energy of 500 GeV in the center-of-momentum system are listed in Table~\ref{tab:backgrounds}.

\begin{table}
    \centering
    \begin{tabular}{|l|l|l|}
        \hline
        Process & cross-section (pb) & \# events $(10^{5} / 2000 fb^{-1}$)\\
        \hline
        4 fermion & $4.1 \times 10^{2}$ & 8.3 \\
        $WW \rightarrow \nu \mu \nu \mu$ & $3.1\times 10^{2}$ & 6.3 \\
        $ZZ \rightarrow \nu \nu \mu\mu$ & $2.7\times 10^{-3}$ & $5.4 \times 10^{-5}$\\
        $Z \rightarrow \mu \mu$ & $1.1 \times 10^{3}$ & 21.4\\
        $Z \rightarrow \tau \tau$ & $1.1 \times 10^{3}$ & 21.0\\
        \hline
        invisible + $H \rightarrow \mu \mu$ & $2.4 \times 10^{-3}$ & $48\times 10^{-5}$ \\
        \hline
    \end{tabular}
    \caption{Considered processes and their cross-sections as obtained from PYTHIA 6.4}
    \label{tab:backgrounds}
\end{table}

\section{Event and Candidate Selection}
Events were selected by requiring exactly two muons. We assume a muon identification efficiency of 100\%, leading to this cut being 98.7\% efficient on signal, while rejecting 67.0 \% of the considered background events.

\subsection{Signal Candidates}
After pre-selecting events by requiring a pair of identified muons, signal candidates are selected by applying a loose cut on the invariant mass of the muon pair. In order to avoid introducing systematic dependencies, the cut on the invariant mass is very wide around the nominal mass of the Higgs boson of $120 \pm 0.5$ GeV. Additional cuts on the visible energy in the event, the oblateness, and the acoplanarity result in a sample consisting of 25 signal events and 8891 background events. The cuts are listed in Table~\ref{tab:cutflow}.
\begin{table}
    \centering
    \begin{tabular}{|l|c|c|}
        \hline
        Cut & Signal & Background \\
        & efficiency & efficiency \\
        \hline
        100 GeV $<$ di-muon mass $<$ 140 GeV & 95.4\% & 4.1\% \\
        130 GeV $<$ visible energy $<$ 260 GeV & 92.2\% & 44.8\% \\
        0 $<$ acoplanarity $<$ 0.5 & 77.7\% & 62.3\% \\
        oblateness $>$ 0 & 75.9\% & 33.8\%\\
        \hline
    \end{tabular}
    \caption{Efficiencies of the signal selection cuts}
    \label{tab:cutflow}
\end{table}

\section{Signal Extraction}
\begin{wrapfigure}{l}{0.45\columnwidth}
    \centerline{\includegraphics[width=.45\columnwidth]{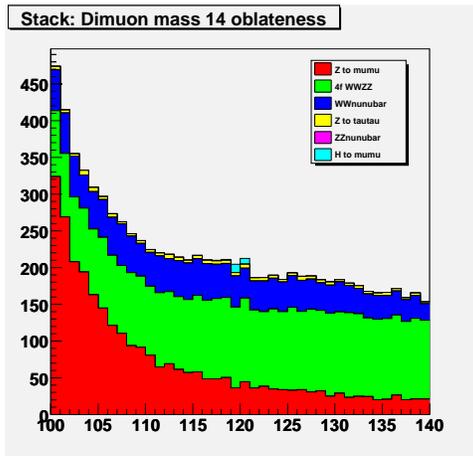}}
    \caption{Stack of the distributions of the di-muon invariant mass after cuts for each of the samples}
    \label{fig:massPlot}
\end{wrapfigure}

Since the other variables available to us do not exhibit clear distinctive features between signal and background, a square cut would reduce the signal efficiency to unacceptable levels. With the help of multivariate classifiers, we take advantage of the full statistical information available to us. Splitting the data sample into a training and a validation set, and providing the classifier with a number of variables in the training set results then in the optimal case in maximally separated signal and background classes. In this analysis, we found that boosted decision trees (also called ``random forest'')~\cite{decisionTrees} exhibit the best performance of the available classifiers and indeed outperform the more commonly used neural nets. With the following set of input variables, we achieve a classification of signal events with a statistical significance of 1.85 sigma. Figure~\ref{fig:massPlot} shows a stack of the distributions of the di-muon invariant mass for each of the samples after cuts.

\begin{itemize}
\item opening angle of the muon pair
\item missing momentum
\item $\cos(\theta)$ for each muon
\item energy for each muon
\item transverse momentum for each muon
\item whether the muons traversed barrel or endcap
\item ratio of the muon energies
\item ratio of $p_T$ of the muons
\end{itemize}
\newpage
\section{Summary and Outlook}
We have presented the status of the analysis of the decay $e^{+} e^{-} \rightarrow \text{ invisible } + H(\rightarrow \mu \mu)$ in the framework of a fast simulation of the sid01 detector. We expect to improve upon the achieved signal significance of 1.85 sigma by including a larger sample and by adding a maximum likelihood fit for the signal extraction.
 
This study was carried out at a collision energy of 500 GeV. Events mediated by the s-channel diagram~\ref{Fig:ZH} are easier to separate from background, because the missing mass exhibits a peak at the nominal Z mass. The fraction of these events in the signal sample is 16.2\% at 500 GeV. For the LOI benchmarking effort, the analysis will be repeated at a collision energy of 250 GeV, reducing the relative fraction of t-channel events in the signal sample to 16.0\%.

\section{Acknowledgments}

The authors would like to expresses thanks to the org.lcsim development team for producing a stable simulation and reconstruction framework for future linear colliders.

\begin{footnotesize}

\end{footnotesize}

\begin{thebibliography}{99}
\bibitem{url} Presentation: \\
{\tiny \verb$http://ilcagenda.linearcollider.org/contributionDisplay.py?contribId=397&sessionId=16&confId=2628$}
\bibitem{pythia} Torbjorn Sjostrand, Stephen Mrenna, Peter Skands, JHEP 0605:026 (2006).
\bibitem{sid} The SiD detector outline document \texttt{http://hep.uchicago.edu/~oreglia/siddod.pdf}
\bibitem{tmva} Andreas Hocker {\it et~al.}, PoS {\bf ACAT} 040 (2007)
\bibitem{decisionTrees} Leo Breiman, Machine Learning 45 (1), 5-32 (2001)
\end{thebibliography}
\end{document}